\documentclass[11pt,aps,preprint]{revtex4}
\usepackage[active]{srcltx}
\usepackage[utf8]{inputenc}
\usepackage{latexsym}
\usepackage{amsmath}
\usepackage{graphicx}

\begin{document}

\title{On Higher Spatial Derivative Field Theories}

\author{Pedro R. S. Gomes}
\email{pedrorsg@fma.if.usp.br}
\affiliation{Instituto de F\'\i sica, Universidade de S\~ao Paulo\\
Caixa Postal 66318, 05315-970, S\~ao Paulo, SP, Brazil}%

\author{M. Gomes}
\email{mgomes@fma.if.usp.br}
\affiliation{Instituto de F\'\i sica, Universidade de S\~ao Paulo\\
Caixa Postal 66318, 05315-970, S\~ao Paulo, SP, Brazil}%


\begin{abstract}

In this work, we employ renormalization group methods to study the general behavior of field theories possessing anisotropic scaling in the spacetime variables. The Lorentz breaking symmetry that accompanies these models are either soft, if no higher spatial derivative is present, or it may have a more complex structure if higher spatial derivatives are also included. Both situations are discussed in models with only scalar fields and also in models with fermions as a Yukawa like model.
\end{abstract}
\maketitle


\section{Introduction}
Quantum Field Theories with anisotropic spacetime scaling have been recently considered in the literature  in the search for viable frameworks for
quantum gravity \cite{Horava} and also for the completion of nonrenormalizable effective field  theories \cite{Anselmi}. The fundamental assumption of theses approaches is that time and space behave
differently under a general scaling: $x^i \rightarrow b x^i$ whereas $ t\rightarrow b^{z}t $. With the "critical exponent" $ z $ conveniently chosen, renormalizability can be achieved  without the undesirable introduction of ghost degrees of freedom which would be  present if higher time derivatives were also included \cite{Antoniadis}. Various investigations on the properties and applications of these approaches may be found in the literature \cite{Sumit}.
However, the anisotropic scaling inevitably encompasses a breaking of Lorentz invariance so that a basic question here  is if  and in what circumstances this symmetry is restored. Although an intuitive argument
says that the restoration may take place at low momenta, a general study of the renormalization group flows for  scalar models indicates that in reality it requires a careful analysis \cite{Iengo} (see also \cite{Iengo1}).

Concerning the Lorentz symmetry breaking, different situations may arise from modifications on the kinetic part of a given Lagrangian that keep the dependence on the time derivative unchanged:

1. The coefficient of the  spatial derivative is modified but higher derivatives are not present.  In  models  for a single field this is physically
  innocuous since it  may be adjusted to any finite value by a mere  change of dimensional units; no breaking of Lorentz symmetry really occurs.  However, in models with more than one field  the modification may have physical implications with radiative corrections which propagate with different velocities such  that the breaking of Lorentz symmetry may increase with the energy. In \cite{Iengo} this possibility was  shown to occur  in a model of fermions and bosons interacting via a trilinear coupling  in four dimensions. As we shall see, the same happens  in a model of two scalar fields
  coupled through trilinear interaction terms in six dimensions. For completeness, we also discuss the case of a Yukawa like model. Whenever the Lorentz symmetry may be restored we say that the breaking is soft.

   2. Higher spatial derivatives are introduced, i.e., we have a truly anisotropic model. In this case,  the coefficient of some of the  higher derivatives  generally changes by effect of the radiative corrections in such way that the breaking of Lorentz symmetry is either soft in the infrared or becomes stronger by decreasing the  energy. It must be stressed here that by soft in this anisotropic situation only means that the effective coefficient of a higher spatial derivative term decreases; in general it can not vanish as the model may become nonrenormalizable.
    In this situation, we shall consider the possibility of the Lorentz symmetry restoration in an approximated sense.

In this work we will pursue these studies focusing some aspects of the renormalization of anisotropic field theories.
With the ultraviolet improved  free propagators, usual Lorentz symmetric models that are renormalizable in a certain dimension, will become  renormalizable in a higher
dimension. Thus with $z=2$, the $\varphi^{3}$ and $\varphi^{4}$ models that are renormalizable in $6$ and $4$ space-time dimensions will become renormalizable in $10$ and $6$ spatial  dimensions, respectively. It should be noticed that although closed forms for the Feynman amplitudes are in general  unfeasible, it is possible to calculate  the renormalization constants and to determine  the  flows of relevant coupling constants.  This may be done by an  application of the BPHZ renormalization theorem to dimensionally regularized integrals as  it will be described shortly.
As an application, we consider the $ \varphi^3  $ model in ten dimensions and with $z=2$ which is asymptotically free. Due to this property, it is not possible to obtain reliable results in the small momenta region. Nevertheless, this model furnishes a simpler setting to expose our methods and discuss some properties of anisotropic field theories as unitarity. Afterwards, we analyze the $\varphi^{4}$ model in six space dimensions which presents an increased degree of complexity as the calculations of the renormalization constants involve a two-loop diagram. In spite of that,
for a special configuration of the initial values of the effective  parameters, we found an explicit solution for the renormalization group equations.

We also consider a model of fermionic and bosonic
fields interacting via a Yukawa like Lagrangian in six space dimension with $ z=2 $. As we shall see, differently from the $ \varphi^3 $ model in $ 10 $ spatial dimensions, this model is infrared stable at the origin  and so it seems to be a good candidate for testing the recuperation of Lorentz symmetry at low momenta. Indeed,  for the four dimensional model and without the higher derivative term we will demonstrate that Lorentz symmetry is restored for  small momenta if initially it was broken by assuming different light velocities for the boson and fermionic components.  When higher spatial derivatives are introduced, we performed a numerical analysis, which is required by the complexity of the renormalization group equations  which prevents the existence closed analytical solutions. In this last situation, we found that the Lorentz symmetry breaking parameters also decrease in the infrared region.

Our work is organized as follows.  In section II, we consider a six dimensional model of two scalar fields interacting through trilinear couplings as a tool to study the soft breaking of Lorentz symmetry; in this preliminary study no higher derivatives are present. In this context, we also analyze the soft breaking of the Lorentz symmetry in a Yukawa like model.  In section III, we present some general remarks and the formalism we shall use in this work. {A subsection is devoted to the unitarity problem in higher spatial derivative models. Sections IV and V are dedicated to the analysis of renormalizable versions of the $\varphi^{4}$ and Yukawa models which are infrared stables.
 A summary and additional remarks are presented in the Conclusions. Two appendices to study integrals needed in our studies
 and provide additional details on the computation of the renormalization group parameters are included.

\section{Soft breaking of Lorentz symmetry}

Before considering truly anisotropic models which contains higher derivatives here we will analyze the soft breaking of Lorentz symmetry in two situations, namely, a six dimensional model with two scalar fields and a Yukawa like model in four dimensions. Let us begin by considering the purely bosonic model
 described by the Lagrangian density
\begin{eqnarray}
{\cal L} &=& \frac{1}{2}\partial_{0}\varphi \partial_{0}\varphi -\frac{b_{\varphi}^{2}}{2}\partial_{i}\varphi \partial_{i}\varphi-\frac{m_{\varphi}^{2}}{2}\varphi^{2}+ \frac{1}{2}\partial_{0}\phi \partial_{0}\phi -\frac{b_{\phi}^{2}}{2}\partial_{i}\phi \partial_{i}\phi-\frac{m_{\phi}^{2}}{2}\phi^{2}\nonumber\\
&& -\frac{\lambda_{1}}{3!}\varphi^{3}-\frac{\lambda_{2}}{3!}\phi^{3}- \frac{\lambda_{3}}{2}\varphi^{2}\phi-\frac{\lambda_{4}}{2}\phi^{2}\varphi\label{I}.
\end{eqnarray}
 The free propagators for the $\varphi$ and $\phi$ fields derived from this Lagrangian,
\begin{equation}
\Delta_{\varphi}(p)= \frac{i}{k_{0}^{2}-b_{\varphi}^{2}{\bf p}^{2}- m_{\varphi}^{2}+i\epsilon}\qquad \mbox{and} \qquad \Delta_{\phi}(p)= \frac{i}{k_{0}^{2}-b_{\phi}^{2}{\bf p}^{2}- m_{\phi}^{2}+i\epsilon},
\end{equation}
will be represented by dashed and dotted lines, respectively. With this graphical notation, the one-loop contributions for the two point vertex functions $\Gamma_{\varphi}^{(2)}$ and $\Gamma_{\phi}^{(2)}$ are drawn in Fig. \ref{Phi3am}.  Notice that  the corresponding analytic expressions, which are supposed to be  regularized by taking the model to $D=6-\epsilon$ dimensions, are exchanged by the replacements $(\varphi, \lambda_{1}, \lambda_{4},\lambda_{3})\leftrightarrow (\phi,\lambda_{2},\lambda_{3},\lambda_{4}) $.   The renormalization group flows of the parameters in (\ref{I}) are fixed by the  introduction
of dimensionless coupling constants $\lambda_{i}\rightarrow \mu^{\frac{\epsilon}{2}}\lambda_{i}$ and the computation of the pole part (PP) of the relevant diagrams.

\begin{figure}[!h]
\centering
\includegraphics[scale=0.8]{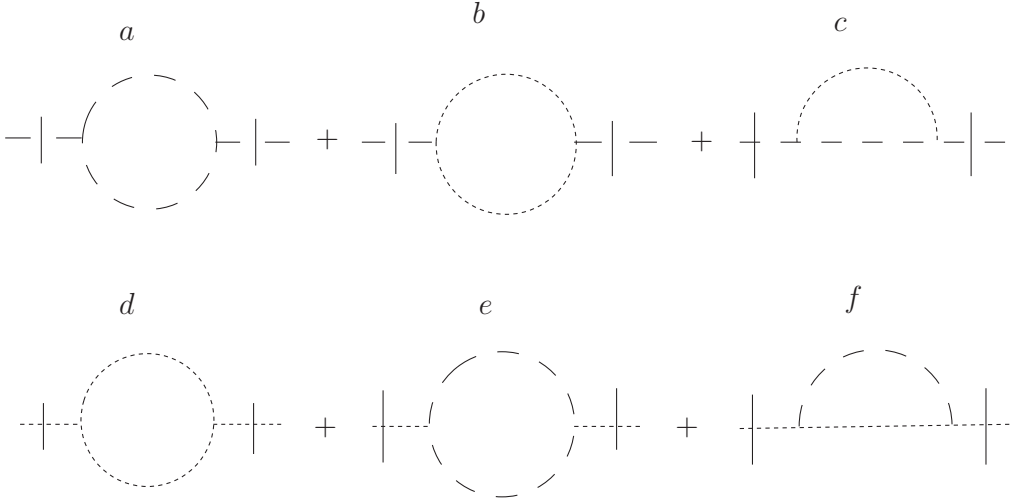}
\caption{One-loop corrections to the two point vertex functions of the $\varphi$ (first row) and $\phi$ (second row) fields.}
\label{Phi3am}
\end{figure}

Following standard procedures, we found that the dependence on the external momentum of the
 pole part of the diagrams in Fig. \ref{Phi3am} are

\begin{eqnarray}
PP[Fig\ref{Phi3am}a]&=& \frac{i(p_{0}^{2}-b_{\varphi}^{2}{\bf p}^{2})\lambda_{1}^{2}}{12 (4\pi)^{3}b^{5}_{\varphi}}\frac{2}{\epsilon},\quad \!PP[Fig\ref{Phi3am}b]= \frac{i({p_{0}^{2}-b_{\phi}^{2}\bf p}^{2})\lambda_{4}^{2}}{12 (4\pi)^{3}b^{5}_{\phi}}\frac{2}{\epsilon},\nonumber\\
PP[Fig\ref{Phi3am}c]&=& \frac{i\lambda_{3}^{2}}{(4\pi)^3b^{5}_{\phi}}\left[\frac{1}{6}p_{0}^{2}-\left(\frac{b_{\varphi}^{2}}{2}-\frac{1}{3}\frac{b_{\varphi}^{4}}{b_{\phi}^{2}}\right) {\bf p}^{2}\right]\frac{1}{\epsilon},\nonumber\\
 PP[Fig\ref{Phi3am}f]&=& \frac{i\lambda_{4}^{2}}{(4\pi)^3b^{5}_{\varphi}}\left[\frac{1}{6}p_{0}^{2}-\left(\frac{b_{\phi}^{2}}{2}-\frac{1}{3}\frac{b_{\phi}^{4}}{b_{\varphi}^{2}}\right) {\bf p}^{2}\right]\frac{1}{\epsilon},
\end{eqnarray}
whereas $PP[Fig\ref{Phi3am}d]$ and $PP[Fig\ref{Phi3am}e]$ are obtained from $PP[Fig\ref{Phi3am}b]$ and $PP[Fig\ref{Phi3am}a]$ by replacing $\lambda_{4}$ by $\lambda_{2}$ and $\lambda_{1}$ by $\lambda_ {3}$, respectively.
Using these results we may obtain the beta functions associated with the parameters $b_{\phi}^{2}$ and $b_{\varphi}^{2}$:
\begin{eqnarray}
\beta_{b_{\varphi}^{2}}&=& \frac{1}{3(4\pi)^{3}b_{\varphi}^{5}}(b_{\varphi}^{2}-b_{\phi}^{2})(\frac{\lambda_{4}^{2}}{2}+\frac{b_{\varphi}^{2}}{b_{\phi}^{2}}\lambda_{3}^{2}),\\
\beta_{b_{\phi}^{2}}&=& -\frac{1}{3(4\pi)^{3}b_{\phi}^{5}}(b_{\varphi}^{2}-b_{\phi}^{2})(\frac{\lambda_{3}^{2}}{2}+\frac{b_{\phi}^{2}}{b_{\varphi}^{2}}\lambda_{4}^{2}).
\end{eqnarray}

\noindent
The basic reason why these expressions do not depend neither on $\lambda_{1}$ or on $\lambda_{2}$ is that the would be contributions are the same as in the case where there is only one
field, either $\varphi$ or $\phi$. As we mentioned in the Introduction, in those cases the parameter $b_{\varphi}$ or $b_{\phi}$ does not receive radiative contributions what implies that the corresponding
$\beta$ function vanishes.

Notice that $b_{\phi}=b_{\varphi}\equiv b$ is a  renormalization group fixed point where the Lorentz symmetry holds. Now, if $b_{\varphi}> b_{\phi}$ then, by lowering the energy,
$b_{\varphi}$ decreases and $b_{\phi}$ increases and the fixed point is infrared stable, the two parameters approaching the fixed point value. Similarly, if $b_{\phi}> b_{\varphi}$, $b_{\phi}$ decreases and $b_{\varphi}$ increases  by lowering the energy the two parameters again  tending to the fixed point $b$. We conclude that in any case the Lorentz symmetry is restored in the low energy regime.

We will also determine the pole part  of the three point vertex functions. Observe that in this case any given diagram either has all three internal lines of the same type or
has two equal lines but of type different from the remaining one. In the last situation the integrals to be computed have the general form
\begin{equation}
I\equiv (-i\lambda)^{3} \int \frac{d^{D} k}{(2\pi)^{D}}\frac{i^{2}}{[k_{0}^{2}-A^{2}{\bf k}^{2}-m^{2}_{A}]^{2}}\frac{i}{k_{0}^{2}-B^{2}{\bf k}^{2}-m^{2}_{B}},
\end{equation}
where $\lambda^{3}$ symbolically represents the factor associated with the coupling constants and either $(A,B)=(b_{\varphi}, b_{\phi})$ or $(A,B)=(b_{\phi},b_{\varphi})$. A straightforward calculation yields then the following pole part:
\begin{equation}
PP(I)= -\frac{i\lambda^{3}}{(4\pi)^{3}}\frac{2(2A+B)}{3A^{3}B(A+B)^{2}}\frac{1}{\epsilon}.
\end{equation}

The above result allows us to  calculate the beta functions associated with the various coupling constants. For example, we obtain
\begin{equation}
\beta_{\lambda_{1}}=\frac{1}{(4\pi)^{3}b_{\phi}^{5}}\left[-\frac{3}{4}\frac{b_{\phi}^{5}}{b_{\varphi}^{5}}\lambda_{1}^{3}-\lambda_{4}^{3}-\frac43\frac{b_{\phi}^{2}(2b_{\phi}+b_{\varphi})}{b_{\varphi}(b_{\phi}+b_{\varphi})^{2}}\lambda_{3}^{2}\lambda_{4}+\left(\frac14-\frac43\frac{b_{\phi}^{4}(2b_{\varphi}+b_{\phi})}{b_{\varphi}^{3}(b_{\phi}+b_{\varphi})^{2}}\right)\lambda_{3}^{2}\lambda_{1}+\frac{1}{4}\lambda_{4}^{2}\lambda_{1}\right].
\end{equation}

Let us now turn our attention to a Yukawa like model specified by
\begin{eqnarray}
\mathcal{L}&=&\frac12 \partial_{0}\varphi \partial_{0}\varphi-
\frac{b_{\varphi}^2}{2} \partial_{i}\varphi \partial_{i}\varphi-\frac{m^2}{2}\varphi^2
+\bar{\psi}(i\gamma^0\partial_0+i b_{\psi}\gamma^i\partial_i-M)\psi\nonumber\\
&+&ig\bar{\psi}\Gamma^5\psi\varphi  -\frac{\lambda}{4!}\varphi^4.
\label{q2}
\end{eqnarray}
In this case we find
\begin{equation}
\beta_{b_{\varphi}^2}=\frac{1}{4\pi^2}\frac{(b_{\varphi}+b_{\psi})}{b_{\psi}^3}(b_{\varphi}-b_{\psi})g^2
\label{q3}
\end{equation}
and

\begin{equation}
\beta_{b_{\psi}}=-\frac{1}{6\pi^2}\frac{1}{b_{\varphi}(b_{\varphi}+b_{\psi})^2}(b_{\varphi}-b_{\psi})g^2,
\label{q4}
\end{equation}

\noindent
which, analogously to the purely bosonic model, shows that $b_{\varphi}=b_{\psi}$ is  also an infrared renormalization fixed point where Lorentz symmetry holds.

\section{Anisotropic scaling: General considerations}\label{example}

The anisotropic field theories with bosonic and fermionic components that we will consider have the generic form
\begin{equation}
{\cal L}= {\cal L}_{0}+ {\cal L}_{int},
\end{equation}
where the free part, ${\cal L}_{0} $, is given by
\begin{equation}
{\cal L}_{0}= \frac12\partial_0 \varphi\partial_0 \varphi+\frac12\sum_{s=1}^{z}\alpha_{s}\partial_{i_1}\ldots\partial_{i_s}\varphi\,\partial_{i_1}\ldots\partial_{i_s}\varphi -\frac{m^{2}}2\varphi^2,
\end{equation}
for bosonic fields and
\begin{equation}
{\cal L}_{0}=  \overline \psi i\gamma_{0}\partial_{0}\psi +\sum_{s=1}^{z}\beta_{s}\overline \psi (i{\bf \gamma}^{i}{\bf \partial}_{i})^{s}\psi-M\overline \psi \psi,
\end{equation}
 for fermionic fields. In these expressions each latin index runs from $ 1  $ to $ d $, the spatial dimension of the model and the effective signs of the $\alpha$'s and $\beta$'s have to be chosen so that the energy  associated with ${\cal L}_{0}$  is positive. $ {\cal L}_{int} $ describes the interaction between these fields. Notice that $ z $ designates the highest degree of the spatial derivatives and we have included terms with less derivatives as they may be necessary in the renormalization process. As each power of $ x_0 $ scales as $ z  $ powers of $ x_i $, the  effective dimension of the Lagrangian is $ z+d $. Then, by taking $ \alpha_z$ and $ \beta_z $ to be dimensionless, we find that effective dimensions of $ \varphi $ and $ \psi $ are respectively

\begin{equation}
\mbox{Dim}[\varphi]= \frac{d-z}2\qquad \mbox{and}\qquad\mbox{Dim}[\psi]= \frac{d}2.
\end{equation}

We will be dealing with Feynman amplitudes of the form
\begin{equation}
\int \prod_{i=1}^{L} dk_i I_G(k,p,m)\label{1a},
\end{equation}
where $L$ is the number of loops, $ k= (k_1,\ldots, k_L)^{} $ and $ p=(p_1,\ldots,p_N) $ are, respectively, the loop momenta and external momenta associated to a generic  proper (i.e. 1PI) Feynman
diagram $ G $. The unsubtracted amplitude $ I_G $ is a product of anisotropic propagators and monomials in the momenta of the lines joining at the vertices of $ G $,
\begin{equation}
I_G(k,p,m)= \prod_{a} P_{a}(k, p) \prod_{abc}\Delta_{F}(l_{abc}),
\end{equation}
with
\begin{equation}
\Delta_{F}(l)= \frac{P(l)}{l_{0}^{2}+\sum_{i=1}^{z}b_i {\bf l}^{2i}- m^{2}+i\epsilon},
\end{equation}
where $ l=(l_{0},\bf l )$ is the momentum flowing through a line of $ G $,  the $ b_i $ are simple functions of the $ \alpha $'s , $ \beta $'s  and $M$ ($b_i= \alpha_{i}$ for a bosonic propagator and  $m=M$ for a fermionic one), and $ P(l) =i$ for a bosonic propagator and a polynomial of first degree in $ l^{0} $ and of $ z$ degree in $ \bf l $, for a fermionic propagator.
The degree of superficial divergence of $ G $ is fixed
by power counting  in which each power of the time like component of a vector counts as the $ z $ power of its spatial like component. This gives
\begin{equation}
d(G)= (d+z)L-2z n_B -zn_F + \sum_a D_a,
\end{equation}
where $ n_B $ and $ n_F $ are the  number of bosonic and fermionic internal lines of $ G $; $D_{a}$ is the degree of  the monomial $ P_{a}(k,p) $ assigned to the vertex $ V_{a} $. Now, if $ G $ has $ V $ vertices, $ L=n_B+n_F -V +1 $ and therefore
 \begin{equation}
d(G)= d+ z -(z-d)n_B+dn_F+ \sum_a(D_a -d-z).
\end{equation}
We also have the topological identities
\begin{equation}
2n_B+N_B=\sum_{a}\nu^{B}_{a}\qquad\mbox{and}\qquad 2n_F+ N_F = \sum_{a}\nu^{F}_{a},
\end{equation}
where $N_{B}$ and $N_{F}$ are the number of external bosonic and fermionic lines, $\nu^{B}_{a} $ and $ \nu^{F}_{a} $ are the number of bosonic and fermionic lines joining at the vertex $ V_a $. Using these relations, we obtain
\begin{equation}
d(G)=d+z -\mbox{Dim}[\varphi] N_B-\mbox{Dim}[\psi] N_F - \sum_{a}(d+z - \mbox{Dim}[V_a])\label{1},
\end{equation}
where Dim[$V_a $]= $D_{a}$+Dim[$\varphi$]$\nu^{B}_{a}$+Dim[$\psi$]$\nu^{F}_{a}$ is the "canonical" operator dimension of the term in $ {\cal L}_{int} $ associated with $ V_{a} $. As usual, we classify a given vertex $ V_a $ as being non-renormalizable, renormalizable or super-renormalizable according it has dimension greater, equal or less than
$ z+d $. Thus, a purely fermionic theory with a quartic non-derivative self-interaction is renormalizable if $ z=d $ and super-renormalizable if $ z>d $. A renormalized amplitude associated with the graph $ G $ may be obtained by applying subtraction operators arranged according Zimmermann' s forest formula \cite{Zimmermann}.
For the special case in which $ G $ is primitively divergent (i.e., without divergent subgraphs), the integral in
(\ref{1a}) can be made finite by  replacing $ I(G) $ by
\begin{equation}
R(G) = (1-t^{d(G)})I_{G}=I_{G}-\sum_{s=0}^{[\frac{d(G)}{z}]}\frac{p_{0}^{s}}{s!}\frac{\partial^{s}{\phantom a}}{\partial p_{0}^{s}}\sum_{n=0}^{d(G)-sz}\frac{{ p}_{i_{1}}\ldots{ p}_{i_{n}}}{n!}\frac{\partial\phantom a}{\partial{ p}_{i_{1}}}\ldots\frac{\partial\phantom a}{\partial{ p}_{i_{n}}}I_{G},\label{2}
\end{equation}
where $[x]$ is the greatest integer less than or equal to $ x $, $ p_{0}^{s} $ symbolically stands for the product of $ s $ time-like components of an independent  set of external momenta; ${ p}_{i}$ denotes the i-th space-like momentum (with the index of the component implicit) and all derivatives are computed at zero external momenta.
Actually, in our one-loop calculations performed in  the next sections we will use the above result just to unveil most easily the pole part of dimensionally regularized amplitudes; afterwards we apply our renormalization  prescription which
consists in removing these pole parts, what is usually called MS subtraction scheme. Besides, throughout   this work we will  take the critical exponent $z$ to be  two.

As in the usual isotropic situation, massless theories requires a special consideration. In these cases it is better to use modified Taylor operators so that the last subtraction is performed replacing $m$ by an auxiliary mass parameter $\mu$ which plays the role of a renormalization point.

Using this scheme, the BPHZ normal product algorithm can be extended to the present situation. Thus,
if $ {\cal O} $ is  a formal product of the basic fields  and their derivatives, a normal product of degree $ \delta $, $ N_{\delta}[{\cal O}] $,
is  defined in the usual way \cite{Lowenstein}. Notice that $ \delta= {\rm Dim}[{\cal O}] + c $, where $c$ is a non-negative integer and the dimension of $ {\cal O} $, ${\rm Dim}[{\cal O}]$, is computed counting $ z $ for each "time" derivative, $ 1 $ for each spatial derivative and the dimensions
of the basics fields as fixed before. As in the isotropic situation, these normal products satisfy  a number of convenient properties
which allows a systematic way for deriving Ward identities and computing their anomalies. In particular, it should be noticed that, inside a Green function,
\begin{equation}
\partial_{i}N_{\delta}[{\cal O}]= N_{\delta+1}[\partial_{i}{\cal O}],\qquad \partial_{0}N_{\delta}[{\cal O}]= N_{\delta+z}[\partial_{0}{\cal O}].
\end{equation}

\subsection*{A simple example}
As a simple example of the methods exposed in the previous section,  we will examine a renormalizable anisotropic $ \varphi^3 $ model which was first considered in \cite{Iengo}.
 Renormalizability requires $ {\rm Dim} [\varphi^3] =d+z  $,  which for $ z=2 $ fixes $ d=10 $. In this situation, the degree of superficial divergence
 of a proper graph $ G $ is given by
\begin{equation}
d(G)=12-4 N,
\end{equation}
where $ N $ is the number of external lines of $ G $. Taking this into consideration, the Lagrangian
including counterterms to cancel the pure pole part is
\begin{eqnarray}
\mathcal{L}&=&\frac12 \partial_{0}\varphi \partial_{0}\varphi-\frac{b^2}{2}\partial_{i}\varphi \partial_{i}\varphi
-\frac{m^2}{2}\varphi^2 -
\frac{a^2}{2}\varphi \Delta^2\varphi+
\frac12 (Z_{\varphi}-1)(\partial_{0}\varphi \partial_{0}\varphi-m^2\varphi^2)\nonumber\\&-&
\frac{b^2}{2}(Z_b-1)\partial_{i}\varphi \partial_{i}\varphi-\frac{\delta m^2 Z_{\varphi}}{2}\varphi^2-\frac{a^2}{2} (Z_{a}-1) \varphi\Delta^2\varphi
-\frac{\lambda Z_{\lambda}}{3!}\varphi^3 + c\varphi= {\cal L}_0 + {\cal L}_{int}
\label{2.57},
\end{eqnarray}
where $\Delta$ is the spatial Laplacian  and
\begin{equation}
{\cal L}_0 =\frac12 \partial_{0}\varphi \partial_{0}\varphi-\frac{b^2}{2}\partial_{i}\varphi \partial_{i}\varphi
 - \frac{a^2}{2}\varphi \Delta^2\varphi -\frac{m^2}{2}\varphi^2,
\end{equation}
with the constant $c$  adjusted to eliminate all tadpoles.
Thus, at one-loop order divergent graphs have two or three external lines and are
quartically or logarithmically divergent, respectively.

 Let us analyze each case separately:

I. One-loop correction to the two point function. The regularized amplitude is given by

\begin{equation}
\Sigma(p)=\frac{\lambda^2}{2}\int\frac{dk_0}{2\pi}\frac{d^dk}{(2\pi)^d}\frac{1}{k^2-a^2 ({\bf k}^2)^2 -m^2+i\epsilon}
\frac{1}{(k+p)^2-a^2 [({\bf k}+{\bf p})^2]^2-m^2+i\epsilon},
\label{2.15}
\end{equation}
where $k^2\equiv k_0^2- b^2{\bf k}^2$ and $ d=10-\epsilon $. As mentioned before,  because of the higher power of the spatial momentum in the denominators, $ \Sigma $ does not possess a closed analytic expression.  However, to compute its pole part it is enough to calculate
the action of the Taylor operator $ t^{4} $ on the above integrand. This leads to integrals which, after using the spatial rotational symmetry, have the general form
\begin{equation}
J(x,y,z)\equiv \int \frac{dk_0}{2\pi}\frac{d^dk}{(2\pi)^d}\frac{k_0^x |{\bf k}|^y}
{[k_0^2-b^2{\bf k}^2-a^2({\bf k}^2)^2+m^2]^z},
\label{4.19a}
\end{equation}
where for $ d=10 $ the parameters $ x, y $ and $ z $ are such that  $ 2x+y-4z+12 $ is either equal to $ 0 $, $ 2 $ or $ 4 $. In these
cases, neglecting finite parts, $ J(x,y,z) $ is given by Eq. (\ref{6}) in the Appendix \ref{apendicea}.
With the help of that result, we have:

 a. Term with four derivatives.
Omitting some finite contributions, the term with four spatial derivatives yields
\begin{eqnarray}
&&\left.\frac{{p}_i{p}_j{p}_k{ p}_l}{4!}\frac{\partial^4 \Sigma(p)}{\partial{ p}_i\ldots\partial {p}_l}\right |_{{p}=0}\!\!=\frac{\lambda ({\bf p}^2)^{2}}{2^4}\int \frac{dk_0}{2\pi}\frac{d^dk}{(2\pi)^d}\left( \frac{32 a^4({\bf k}^2)^2}{[\text{den}]^4}+\frac{8 a^2}{[\text{den}]^3}\right. +
\frac{768 a^6({\bf k}^2)^4}{d[\text{den}]^5}+\frac{3 (2)^7 a^4 ({\bf k}^2)^2}{d[\text{den}]^4}
\nonumber\\
&&+\frac{ 3 (2)^{11} a^8({\bf k}^2)^6}{d(d+2)[\text{den}]^6}+\frac{9 (2)^9 a^6 ({\bf k}^2)^4}{d(d+2)[\text{den}]^5}+
\left.\frac{3 (2)^7 a^4 ({\bf k}^2)^2}{d(d+2)[\text{den}]^4}\right),
\end{eqnarray}
where $\text{den}\equiv k^2-m^2-a^2 ({\bf k}^2)^2+i\epsilon$ . Performing the integrals, we find that  their pole part yields the result

\begin{equation}
-i \frac{11}{15 (2)^{18}\pi^5}\frac{\lambda^2}{a^3}\frac{({\bf p}^2)^2}{d-10}.
\end{equation}

b. The term with two derivatives with respect to the components of ${\bf p} $ may be calculated analogously. One finds
\begin{equation}
\left.\frac{{p}_i {p}_j}{2}\frac{\partial^2 \Sigma(p)}{\partial{p}_i\partial {p}_{j}} \right |_{{\bf p}=0}=\frac{\lambda^2{\bf p}^2}2\int \frac{dk_0}{2\pi}\frac{d^dk}{(2\pi)^d}\left(\frac{16 b^2 a^2({\bf k}^2)^2}{d[\text{den}]^4}+\frac{16 a^4({\bf k}^2)^3}{d[\text{den}]^4}+
\frac{b^2}{[\text{den}]^3}+\frac{2 a^2 {\bf k}^2}{[\text{den}]^3}
+\frac{4a^2{\bf k}^2}{d[\text{den}]^3}\right),
\end{equation}
such  that the pole part is
\begin{equation}
-i \frac{5}{3 (2)^{17}\pi^5}\frac{ b^2 \lambda^2}{a^5}\frac{1}{d-10} {\bf p}^2\label{3}.
\end{equation}

c. The term with two derivatives with respect to $ p_0 $ gives
\begin{equation}
\frac{p_{0}^{2}}2\frac{\partial^2 \Sigma(p)}{\partial p_0\partial p_0}\Big{|}_{p=0}=\frac{p_{0}^{2}}2\int \frac{dk_0}{2\pi}\frac{d^dk}{(2\pi)^d}\left (\frac{3}{[\text{den}]^3}+
\frac{4 a^2 ({\bf k}^2)^2}{[\text{den}]^4}\right)\,,
\label{2.41}
\end{equation}
providing the pole part
\begin{equation}
-i \frac{1}{ 3 (2)^{17} \pi^5} \frac{\lambda^2}{a^{5}}  \frac{1}{d-10}\,  p_{0}^{2}.
\label{4}
\end{equation}

d. Similarly, the term without derivatives computed at zero external momentum furnishes the pole part
\begin{equation}
i \frac{1}{(2)^{18}\pi^5}\frac{\lambda^2}{a^{7}}\frac{1}{d-10}\, (-5b^4+4 a^2 m^2).
\label{2.49}
\end{equation}

Collecting the above results we can write

\begin{equation}
\Sigma(p)= i \frac{1}{(2)^{18}\pi^5}\frac{\lambda^2}{d-10}\left ( \frac{4 a^2 m^2-5b^4}{a^7}-\frac23 \frac{p_{0}^{2}}{a^5}-\frac{10\, b^2}3
\frac{{\bf p}^2}{a^5}-\frac{11}{15} \frac{({\bf p}^2)^2}{a^3}\right)+ \mbox{finite terms}.
\end{equation}

II. To complete our computation of the counterterms, we need to  consider the three point function. At one-loop a direct calculation gives that
\begin{equation}
\Gamma^{(3)}= -i\lambda +i \frac{1}{2^{16} \pi^5} \frac{\lambda^3}{a^{5}}  \frac{1}{d-10} +\mbox{finite terms}.
\label{2.54}
\end{equation}

We are now in a position to determine the renormalization group flows of the parameters of the model. In fact, the $ N $ point vertex functions of the model satisfy a 't Hooft-Weinberg renormalization group equation

\begin{equation}
\left[\mu\frac{\partial}{\partial\mu}+\delta\frac{\partial}{\partial m^2}+\beta_{b^{2}}\frac{\partial}{\partial b^2}
+\beta_{a^{2}}\frac{\partial}{\partial a^2}+
\beta_{\lambda}\frac{\partial}{\partial\lambda}-N\gamma\right]\Gamma^{(N)}(p, m^2, b^2,  a^2, \lambda, \mu)=0,
\label{2.65}
\end{equation}

\noindent
where the renormalization scale $ \mu $ was introduced by replacing $ \lambda \rightarrow \mu^{\epsilon/2} \lambda $, where $ \epsilon=10-d $, so that the new coupling constant $ \lambda $ is dimensionless. After removing the pure pole part,  the renormalization group parameters may be then
obtained by replacing into (\ref{2.65}) the renormalized vertex functions

\begin{equation}
\Gamma^{(2)}=i[p_0^2-b^2{\bf p}^2-a^2({\bf p}^2)^2-m^2-\lambda^2 (\text{Finite}_1-\ln\mu\text{Residue}_1)]
\label{2.72}
\end{equation}
and
\begin{equation}
\Gamma^{(3)}=-i\lambda +i \lambda^3\left(\text{Finite}_2-\ln\mu\text{Residue}_2\right),
\label{2.78}
\end{equation}
where $ \text{Finite}_{1,2} $ are the finite parts  and $ \text{Residue}_{1,2} $ the residues at the poles of the corresponding vertex functions. We find
\begin{eqnarray}
\beta_{b^{2}}&=&\frac{ b^2}{2^{16}\pi^5}\frac{\lambda^2}{a^{5}},\qquad \beta_{a^{2}}=\frac{7}{5 (2)^{18}\pi^5}\frac{\lambda^2}{a^{3}}, \qquad \beta_{\lambda}=-\frac{3}{2^{18}\pi^5}\frac{\lambda^3}{a^{5}}, \nonumber\\
\delta & =& \frac{(5b^4-\frac{10}{3}\, m^2 a^2)}{ 2^{18}\pi^5}\frac{\lambda^2}{a^{7}}\qquad \text{and} \qquad \gamma=\frac{1}{ 3 (2)^{18} \pi^5} \frac{\lambda^2}{a^{5}},
\label{2.95}
\end{eqnarray}
which agree with \cite{Iengo} and shows that the model is asymptotically free. Furthermore, the effective mass increases or decreases with $ \mu $ accordingly $ b^4 $ is greater or less than $ 2/3 m^2 a^2 $. By introducing a logarithmic scale, $ t=\ln(\mu/\mu_{0} )$, where $ \mu_0 $ is a reference scale where the parameters in the Lagrangian have been defined, we may evaluate the flows of the effective parameters as follows,
\begin{equation}
\frac{\partial{\overline{a}}^2}{\partial t}=\frac{7}{5 (2)^{18}\pi^5}\frac{\overline{\lambda}^2}{\overline{a}^{3}}, \qquad \text{with}\qquad \overline a(0)= a
\label{2.99}
\end{equation}
and
\begin{equation}
\frac{\partial\overline{\lambda}}{\partial t}=-\frac{3}{2^{18}\pi^5}\frac{\overline{\lambda}^3}{\overline{a}^{5}},\qquad\text{with}\qquad
\overline\lambda(0)=\lambda .
\label{2.100}
\end{equation}
These equations imply  that
\begin{equation}
\overline{\lambda}\frac{\partial^2 \overline \lambda}{\partial t^2}
=\frac{25}{6}\left(\frac{\partial \overline \lambda}{\partial t}\right)^2,
\label{2.102}
\end{equation}
whose general solution is
\begin{equation}
\overline{\lambda}(t)=\frac{c_2}{(19 t+6 c_1)^{\frac{6}{19}}}\, ,
\label{2.103}
\end{equation}
where $c_1$ and $c_2$ are constants  determined by the initial conditions. Using this result,  we integrate the equation for $ \overline{a} $ giving
\begin{equation}
\overline{a}^{5}(t)=\frac{c_2^2}{2^{19}\pi^5}(19 t+6 c_1)^{\frac{7}{19}}
\label{2.104}
\end{equation}
and also
\begin{equation}
c_1=\frac{2^{18}\pi^{5}} {3}\frac{a^{5}}{\lambda^2} \qquad \text{and}\qquad c_2=64 \pi^{\frac{30}{19}}(a)^{\frac{30}{19}}\lambda^{\frac{7}{19}}.
\label{2.105}
\end{equation}

Notice that $ \overline{a} $ decreases with $ t $ vanishing at the critical value $ t_{IR}=-\frac{2^{19}}{19}\frac{a^{5}}{\lambda^2}$. The same happens with the effective parameter $ \overline{b} $ which is given by
\begin{equation}
\overline{b}^2(t)=\frac{b^2}{(6 c_1)^{\frac{8}{19}}}(19 t+6 c_1)^{\frac{8}{19}}\,  .
\label{2.107}
\end{equation}
However, at the same time the effective coupling constant increases tending to infinity as $ t $ approaches $ t_{IR} $. Our perturbative methods are not applicable insofar $ \overline{\lambda} $ is not small and no conclusion can be made concerning the restoration of  Lorentz symmetry at small momenta.

\subsubsection{Unitarity in the anisotropic situation.} Differently from what happens in Lorentz invariant theories   but with higher space and time derivatives, models where only higher spatial derivatives are introduced  preserve unitarity (in this respect see also \cite{Albrecht}). Let us exemplify this by analyzing the Cutkosky rules \cite{diagrammar} for the one-loop contribution
to the two point function of the model (\ref{2.57}). Those rules demand that the diagrammatic relation of Fig. \ref{U1} be obeyed. The analytic expression for the graph on the left hand side is
\begin{equation}
iT=\frac{\lambda^2}{2}\int\frac{dk_0}{2\pi}\frac{d^dk}{(2\pi)^d}\frac{1}{k_0^2-\omega_k^2+i\epsilon}
\frac{1}{(p_0-k_0)^2-\omega_{p-k}^{2}+i\epsilon},
\label{U.1}
\end{equation}
where the only restriction on  $\omega_{k}$ is that it  depends only on the spatial part of the momentum. For the model (\ref{2.57}), $\omega_{k}^{2}=b^{2} {\bf k}^{2}+a^{2}({\bf k}^{2})^{2}+m^{2}$.
By integrating over $ k_{0} $ and using the identity
\begin{equation}
\frac{1}{A\pm i\epsilon}=\text{P}\left(\frac{1}{A}\right)\mp i\pi \delta(A), \label{8}
\end{equation}
where P denote the Cauchy principal value, we get
\begin{equation}
2\,\text{Im}T=\pi\lambda^2\int\frac{d^dk}{(2\pi)^d}
\frac{\delta(p_0-\omega_k-\omega_{p-k})}{2\omega_k\, 2\omega_{p-k}}.\label{7a}
\end{equation}

On the other hand, the expression for the cut diagram on the right of Fig. \ref{U1}  is obtained by the following replacement of  the free propagator
\begin{equation}
\Delta(k)=\frac{i}{k_0^2-\omega_k^2+i\epsilon}\rightarrow \Delta^{+}(k)= 2\pi \theta (k_0)\delta(k_0^2-\omega_k^2)
\label{U.9}
\end{equation}
and by noticing that the vertex factor on the left of the cut is the complex conjugate of the one on the right of it. Therefore, the graph on the right of Fig.\ref{U1} gives
\begin{equation}
\sum |T|^2=\frac{\lambda^2}{2}\int\frac{dk_0}{2\pi}\frac{d^dk}{(2\pi)^d}
2\pi \theta (k_0)\delta(k_0^2-\omega_k^2) 2\pi\theta(p_0-k_0)\delta((p_0-k_0)^2-\omega_{p-k}^2),
\label{U.10}
\end{equation}
which after integrating over $ k_{0} $ gives the same result as in Eq. (\ref{7a}).

\begin{figure}[!h]
\centering
\includegraphics[scale=0.8]{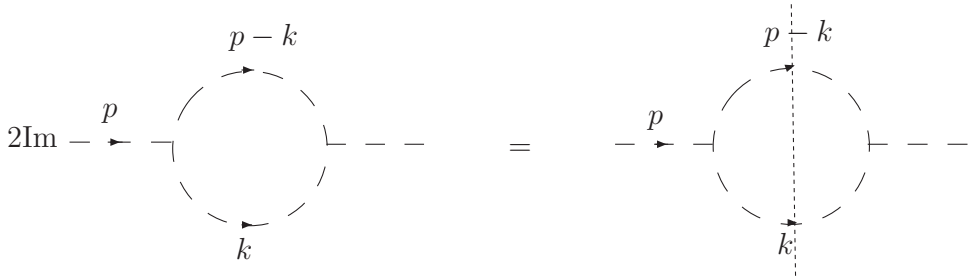}
\caption{Cutkosky rule at one-loop order.}
\label{U1}
\end{figure}

\noindent
\section{The $\varphi^{4}$ model}

The $\varphi^{4}$ model with $z=2$, specified by the Lagrangian density
\begin{equation}
\mathcal{L}=\frac12 \partial_{0}\varphi \partial_{0}\varphi-\frac{b^{2}}{2} \partial_{i}\varphi \partial_{i}\varphi -
\frac{a^{2}}{2}\varphi \Delta^2\varphi-\frac{m^2}{2}\varphi^2-\frac{\lambda}{4!}\varphi^4,
\label{3.1}
\end{equation}
turns out to be renormalizable in $6$ spatial dimensions, the degree of superficial divergence being given by
\begin{equation}
d(G)= 8 -2 N.
\end{equation}

If a BPHZ like scheme is adopted, we find that the
Green functions satisfy the equation of motion

\begin{eqnarray}
<0|{\rm T} N_{8}[\varphi(\partial_{0}^{2}-{b^{2}}\Delta+a^{2}\Delta^{2}+ m^{2})\varphi](x) X|0>&=& -\frac{\lambda}{3!}<0|{\rm T}N_{8}[\varphi^{4}](x) X|0>\nonumber\\
&&-i\sum_{i=1}^{N}\delta(x-x_{i})<0|{\rm T}X|0>,
\end{eqnarray}
where $X= \prod_{i}\varphi(x_{i})$.
This expression may be derived by noting that in momentum space the operator applied on $\varphi$, in the left hand side of the above equation, is equal to $-i$ times the inverse of the free field propagator
\begin{equation}
\Delta_{F}(k)= \frac{i}{k_0^2-b^{2}{\bf k}^2-a^2 ({\bf k}^2)^2-m^2+i\epsilon}.
\end{equation}

To fix the renormalization group parameters, we compute the radiative corrections to the parameters of the model as follows.

The lowest order correction to the two point function comes from the tadpole graph in Fig. \ref{E3a} whose analytic expression is
\begin{equation}
\frac{\lambda}{2}\int\frac{dk_0}{2\pi}\frac{d^dk}{(2\pi)^d}\frac{1}{k_0^2-b^{2}{\bf k}^2-a^2 ({\bf k}^2)^2-m^2+i\epsilon}.
\label{3.3}
\end{equation}
\begin{figure}[!h]
\centering
\includegraphics[scale=0.8]{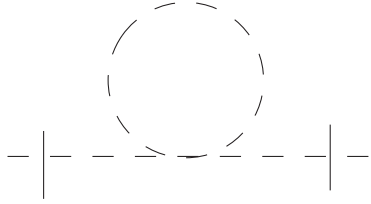}
\caption{Two point function of order $\lambda$.}
\label{E3a}
\end{figure}
\noindent
Since the above integral does not depend on the external  momentum, with the help of (\ref{6}) its divergent part is easily obtained, i.e.,
\begin{equation}
-i\frac{(-3b^2+4 a^{2} m^2)}{2048\pi^3}\frac{\lambda}{a^{5}}\frac{1}{d-6}.
\label{3.5}
\end{equation}

The  one-loop contribution to the coupling constant renormalization  it is also straightforwardly calculated. In fact, we just
need to calculate the pole part of

\begin{equation}
\frac{3\lambda^2}{2}\int\frac{dk_0}{2\pi}\frac{d^dk}{(2\pi)^d}\frac{1}{[k_0^2-b^{2}{\bf k}^2-a^{2} ({\bf k}^2)^2-m^2+i\epsilon]^2},
\label{3.6}
\end{equation}
which, again with the help of (\ref{6}), is found to be
\begin{equation}
-i\frac{3}{512 \pi^3}\frac{\lambda^2}{a^{3}}\frac{1}{d-6}.
\label{3.8}
\end{equation}
To find the lowest order  corrections to the $a$ and $b$ parameters is a much more difficult task since it involves the calculation of a two-loop graph. Actually, only
for $b=0$ and $m=0$ closed analytic expressions exists \cite{Carneiro}. In that case one needs to calculate
\begin{equation}
\frac{i\lambda^2}{3!}\int \frac{dk_0}{2\pi}\frac{d^dk}{(2\pi)^d}\frac{dq_0}{2\pi}\frac{d^dq}{(2\pi)^d}
\frac{1}{k_0^2-a^{2}({\bf k}^2)^2}\frac{1}{q_0^2-a^{2}({\bf q}^2)^2}\frac{1}{(p_0-k_0-q_0)^2-
a^{2}(({\bf p}-{\bf k}-{\bf q})^2)^2},
\label{3.18}
\end{equation}
whose pole part turns out  to be
\begin{equation}
-\frac{i\lambda^2}{288}\frac{\pi^2}{a^4}\left[\frac{1}{a^{2}}\left(1+3\ln\frac34\right)p_0^2+\frac{7}{72}({\bf p}^2)^2\right]
\frac{1}{d-6},
\label{3.19}
\end{equation}
yielding the following beta function
\begin{equation}
\beta_{a^{2}}=\frac{\pi^2}{144}\left(\frac{79}{72}+3\ln\frac34\right)\frac{\lambda^2}{a^4}.
\label{3.20}
\end{equation}
Similarly, the computation of the pole part of the four point vertex function furnishes
\begin{equation}
\beta_{\lambda}^{(2)}=\frac{3}{512 \pi^3}\frac{\lambda^2}{a^{3}},
\label{3.21}
\end{equation}
so that the model is infrared stable.
This result will be used in the next section as the renormalization for the Yukawa model demands the inclusion of the $\varphi^{4}$ self-interaction.

Concerning to the solution of these equations,
there is a particular configuration of the initial values of the
parameters $a$ and $\lambda$, such that the system of differential equations above exhibits a simple analytical solution. In fact,
denoting the constants appearing in above equations by $A\equiv \frac{\pi^2}{144}\left(\frac{79}{72}+3\ln\frac34\right)$
and $B\equiv \frac{3}{512 \pi^3}$, we can get from (\ref{3.20}) and (\ref{3.21}), the following equation for the
effective coupling constant
\begin{equation}
\frac{3A}{2B^2}\left(\frac{\partial \overline\lambda}{\partial t}\right)^3-2\overline\lambda \left(\frac{\partial \overline\lambda}{\partial t}\right)^2
+\overline\lambda^2\left(\frac{\partial^2 \overline\lambda}{\partial t^2}\right)=0,
\label{3.23}
\end{equation}
which admit the solution
\begin{equation}
\overline\lambda(t)=\overline\lambda(0)e^{\frac{3A}{2B^2}t}.
\end{equation}
From (\ref{3.20}), we obtain
\begin{equation}
\overline a(t)=\left[\frac{9A^2}{4B^2}\overline\lambda^2(0)(e^{\frac{3A}{B^2}t}-1)+\overline a^6(0) \right]^{\frac16}.
\end{equation}
But this will be a solution of the system only
under the condition $\overline a^6(0)=\frac{9A^2}{4B^2}\overline\lambda^2(0)$, yielding
\begin{equation}
\overline a(t)=\left(\frac{9A^2}{4B^2}\overline\lambda^2(0)\right)^{\frac16} e^{\frac{A}{2B^2}t}.
\end{equation}
This result show us that the Lorentz symmetry breaking parameter $\overline a$ goes to zero
as $t\rightarrow -\infty$, i.e., in the infrared region.
On dimensional grounds, we see that these solutions would not suffer modifications even in the presence of
the parameters $b$ and $m$, such as in lagrangian (\ref{3.1}). So, we concludes that in this case the Lorentz
symmetry can be approximately recovered in a sufficient low scale of the energy.

\section{A renormalizable Yukawa model with $z=2$}

The analysis of the previous sections indicates that anisotropic models that are infrared stables are the best candidates to show
Lorentz symmetry restoration at small energies.  To further  investigate this possibility
 we consider now a model of boson and fermion fields interacting through the Lagrangian density
\begin{eqnarray}
\mathcal{L}&=&\frac12 \partial_{0}\varphi \partial_{0}\varphi-\frac{b^{2}_{\varphi}}{2} \partial_{i}\varphi \partial_{i}\varphi-\frac{m^2}{2}\varphi^2 -
\frac{a^{2}_{\varphi}}{2}\varphi \Delta^2\varphi\nonumber\\
&+&\bar{\psi}(i\gamma^0\partial_0+i b_{\psi}\gamma^i\partial_i+
a_{\psi}\Delta-M)\psi
+ig\bar{\psi}\Gamma^5\psi\varphi  -\frac{\lambda}{4!}\varphi^4,
\label{4.8}
\end{eqnarray}
where $ \Gamma^{5} $ is the chiral matrix.  In momentum space, the free propagators derived from the above expression are
\begin{equation}
\Delta_F(k)=\frac{i}{k_{0}^{2}-b^{2}_{\varphi} {\bf k}^2-a^{2}_{\varphi} ({\bf k}^2)^2-m^2},
\end{equation}
for the bosonic field $ \varphi $ and
\begin{equation}
S_F(k)=\frac{i}{\gamma^0k^0-b_{\psi} \gamma\cdot{\bf k}-a_{\psi}{\bf k}^2-M},
\label{4.8a}
\end{equation}
for the fermion field.
\begin{figure}[!h]
\centering
\includegraphics[scale=0.8]{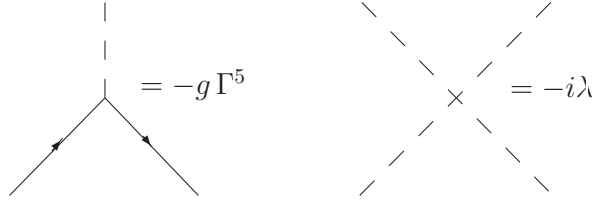}
\caption{Interacting vertices for the Yukawa model.}
\label{FR3}
\end{figure}
 The interaction between the basic fields is given by the non-quadratic part of (\ref{4.8}) so that the corresponding Feynman rules are as depicted in Fig \ref{FR3}. With  the anisotropic scaling exponent $ z=2 $, the model is renormalizable in $ d=6 $ spatial dimensions and one time
dimension. Indeed, in this situation the degree of superficial divergence for a generic graph $ G $ is
\begin{equation}
d(G)=8 -2 N_B-3N_F,
\end{equation}
where $ N_B $ and $ N_F $ are the number of external boson and fermion lines of $ G $.  As the spacetime dimension  is odd, we
 will work with gamma matrices adequate for eight dimensions as to be able to define the chiral matrix as (notice that this assumption is in accord with
 the fact that the effective  dimension  of the underlying spacetime is eight)
\begin{equation}
\Gamma^5\equiv i\prod_{\mu=0}^7\gamma^{\mu}.
\label{5.1}
\end{equation}
The above  choice for the interaction term prevents the induction of counterterms proportional to $ \varphi^{3} $, $ \varphi $ and $ \overline \psi\psi\varphi $ and so is simpler than the other possible renormalizable
 interaction $ \overline \psi \psi \varphi $.

As usual, we define renormalized quantities
through the replacements
\begin{eqnarray}
\varphi\qquad & \rightarrow &\qquad Z_{\varphi}^{1/2} \varphi,\\
\psi\qquad&\rightarrow &\qquad  Z_{\psi}^{1/2} \psi,\\
g\qquad  &\rightarrow & \qquad  \frac{Z_g}{Z_\psi Z_{\varphi}^{1/2}}g,\\
\lambda\qquad &\rightarrow & \qquad  \frac{Z_\lambda}{ Z_{\varphi}^{2}}\lambda
\end{eqnarray}
and also the Lorentz breaking parameters,

\begin{equation}
\{b_{\varphi}^{2},b_\psi,a_{\varphi}^{2}, a_\psi\}\qquad \rightarrow\qquad \{(Z_{b_{\varphi}}/Z_{\varphi}) b_{\varphi}^{2}, (Z_{b_\psi}/Z_\psi) b_\psi, (Z_{a_{\varphi}} /Z_{\varphi}) a_{\varphi}^{2}, (Z_{a_\psi}/Z_\psi )a_\psi\}.
\end{equation}

For the computation of the renormalization constants, we shall use the dimensional reduction scheme \cite{Semenoff} in which all algebraic simplifications are done  with the  Dirac matrices as introduced above   and, afterwards, the   integrals are promoted to
$ d=6-\epsilon $ spatial dimensions. The ambiguities that may be present in such procedure \cite{Collins} only manifest themselves in higher order and do not affect our one-loop computations.
\begin{figure}[!h]
\centering
\includegraphics[scale=0.8]{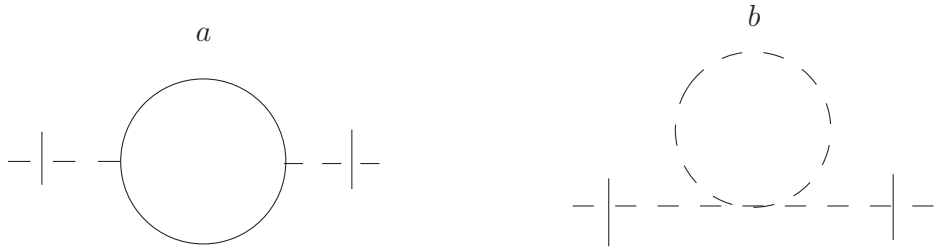}
\caption{ Lowest order contributions to the self-energy of the scalar field.}
\label{E8}
\end{figure}
 Up to one-loop order,  the relevant graphs are shown in Figs.\ref{E8}-\ref{E9}. We begin by considering the order $g^{2}$ radiative correction to the self-energy of the field $\varphi$, Fig. \ref{E8}a, whose unsubtracted analytic form is
\begin{eqnarray}
\Sigma_\varphi(p)&=&g^2\int \frac{dk_0}{2\pi}\frac{d^dk}{(2\pi)^d}
\text{Tr}\left[\Gamma^{5}\frac{1}{\gamma^0k^0-b_{\psi} \gamma\cdot{\bf k}-a_{\psi}{\bf k}^2-M}\right.\nonumber\\
&&\times\left.
\Gamma^{5}\frac{1}{\gamma^0(k^0+p^0)-b_{\psi} \gamma\cdot({\bf k}+{\bf p})-a_{\psi}({\bf k}+{\bf p})^2-M}\right],
\label{4.13}
\end{eqnarray}
where, as we are considering the eight dimensional representation, the gamma matrices  are $ 2^{4} $ dimensional and obey $\text{Tr}\,\gamma^{\mu}=0$ and $\text{Tr}\,\gamma^{\mu}\gamma^{\nu}=2^{4}g^{\mu\nu}$.

As before, for generic momenta the computation of $ \Sigma_\varphi(p) $ is cumbersome and probably unfeasible. However,
to extract its pole part we may proceed as in section \ref{example}.
After discarding some finite contributions, the term of the fourth order derivative with respect to the spatial momenta gives

\begin{eqnarray}
&&g^2 2^{4} \frac{p_ip_jp_kp_l}{4!}\int \frac{dk_0}{2\pi}\frac{d^dk}{(2\pi)^d}
\left[\frac{-k_0^2}{[\text{den}]}\frac{\partial^4}{\partial k_i\partial k_j\partial k_k\partial k_l}\frac{1}{\text{[den]}}
+\frac{a_{\psi}^2{\bf k}^2}{[\text{den}]}\frac{\partial^4}{\partial k_i\partial k_j\partial k_k\partial k_l}
\frac{{\bf k}^2}{[\text{den}]}\right],\nonumber\\
\label{4.17}
\end{eqnarray}
where, for simplicity, we have defined $\text{den}\equiv k_0^2-b_{\psi}^2{\bf k}^2-(a_{\psi}{\bf k}^2+M)^2$. Thus, we have to compute integrals of the form
\begin{equation}
J(x,y,z)\equiv \int \frac{dk_0}{2\pi}\frac{d^dk}{(2\pi)^d}\frac{k_0^x |{\bf k}|^y}
{[k_0^2-b_{\psi}^2{\bf k}^2-(a_{\psi}{\bf k}^2+M)^2]^z},
\label{4.19b}
\end{equation}
which, up to some finite terms,  can be read from the appendix \ref{apendicea}, by sequentially making the replacements $ m^2\rightarrow M^{2} $, $ b^{2} \rightarrow  b^{2}_{\psi}+ 2 M a_{\psi} $ and $ a^{2}\rightarrow a^{2}_{\psi} $.  The pole part at $ d=6 $ gives then
\begin{equation}
-\frac{i}{ (2)^7\pi^3}\frac{g^2}{a_{\psi}}\frac{1}{d-6}({\bf p}^2)^2.
\label{5.9}
\end{equation}
The other terms in the Taylor expansion may be calculated similarly. We find that the pole part  of the self-energy   of the $ \varphi $ field  is
\begin{equation}
-i\left[\frac{(3b_{\psi}^4+12a_{\psi}b_{\psi}^2M+8a_{\psi}^2M^2)}{64\pi^3}\frac{1}{a_{\psi}^{4}}+\frac{1}{32\pi^3}\frac{1}{a_{\psi}^{2}}\,p_0^2+\frac{M}{16\pi^3}\frac{1}{a_{\psi}^{}}\,{\bf p}^2+\frac{1}{3 (2)^{7}\pi^3}({\bf p}^2)^2 \right] \frac{g^2}{a_{\psi}}\frac{1}{d-6}.\label{9}
\end{equation}
The tadpole graph has already been considered in our study of the $\varphi^{4}$ model and we simply quote that  the result is given by Eq. (\ref{3.5}).

Let us now consider the one-loop contribution to the self-energy of the fermion field which corresponds to the graph in  Fig. \ref{E10}. The analytic expression is
\begin{eqnarray}
\Sigma_{\psi}(p)&=&-g^2\int \frac{dk_0}{2\pi}\frac{d^dk}{(2\pi)^d}\,\Gamma^5
\frac{1}{\gamma^0k^0-b_{\psi} \gamma\cdot{\bf k}-a_{\psi}{\bf k}^2-M}\,\Gamma^5\nonumber\\
&&\times
\frac{1}{(p_0-k_0)^2-b_\varphi^2({\bf p}-{\bf k})^2-a_{\varphi}^{2} [({\bf p}-{\bf k})^2]^2-m^2}\nonumber\\
&=&
-g^2\int \frac{dk_0}{2\pi}\frac{d^dk}{(2\pi)^d}
\frac{-\gamma^0k^0+b_{\psi} \gamma\cdot{\bf k}+a_{\psi}{\bf k}^2+M}
{k_0^2-b_{\psi}^2 {\bf k}^2-(a_{\psi}{\bf k}^2+M)^2}\nonumber\\
&&\times
\frac{1}{(p_0-k_0)^2-b_{\varphi}^2({\bf p}-{\bf k})^2-a_{\varphi} ^{2}[({\bf p}-{\bf k})^2]^2-m^2}.
\label{5.22}
\end{eqnarray}
\begin{figure}[!h]
\centering
\includegraphics[scale=0.7]{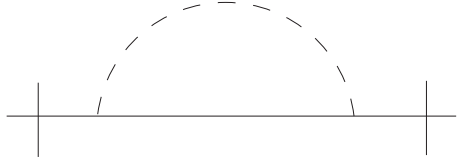}
\caption{Two point fermionic function.}
\label{E10}
\end{figure}
From this we get the pole terms:

a. Term with two  derivatives with respect to the spatial part of the external momentum:
\begin{equation}
  \frac{-i}{3 (2)^{7}\pi^{3}}\frac{a_{\psi}(3 a_{\varphi}+a_{\psi})}{a_{\varphi}(a_{\varphi}+a_{\psi})^{3}}\frac{g^{2}{\bf p}^{2}}{d-6}.\label{12}
  \end{equation}

b. Term with one derivative with respect to $p_{0}$:
\begin{equation}
\frac{-i}{(2)^7\pi^3}\frac{g^2}{a_{\varphi}(a_{\psi}+a_{\varphi})^{2}}\frac{1}{d-6}p_0\gamma^0.\label{13}
\end{equation}

c. Term with one  derivative with respect to  the spatial part of the external momentum:

\begin{equation}
i \frac{1}{3 (2)^7\pi^3}\frac{(2a_{\varphi}+a_{\psi})b_{\psi}}{a_{\varphi}a_{\psi}(a_{\varphi}+a_{\psi})^{2}}\label{14}
\frac{g^{2}}{d-6}{\bf p}\cdot\gamma.
\end{equation}

d. Term without derivatives:
\begin{equation}
-\frac{1}{2^8\pi^3}\frac{g^{2}}{a_{\varphi}^{3}a_{\psi}^{2}}\frac{[a^{2}_{\psi}(2a_{\varphi}+a_{\psi})b_{\varphi}^{2}+a_{\varphi}^{2}(a_{\varphi}+2a_{\psi})b_{\psi}^{2}+
2 a_{\varphi}^{2}a_{\psi}^{2}M]}{(a_{\psi}+a_{\varphi})^{2}}\frac{1}{d-6}.\label{15}
\end{equation}

\begin{figure}[!h]
\centering
\includegraphics[scale=0.7]{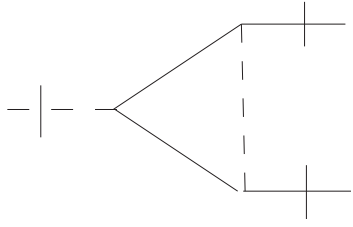}
\caption{Three point vertex function.}
\label{E11}
\end{figure}

 The one-loop contribution to the three point vertex function, Fig. \ref{E11}, furnishes

\begin{equation}
-\frac{1}{2^7\pi^3}\frac{1}{a_{\varphi}a_{\psi}(a_{\varphi}+a_{\psi})}g^3\Gamma^5\frac{1}{d-6},
\label{16}
\end{equation}

whereas for the four point vertex function of the bosonic field shown in Fig. \ref{E9} we obtain
\begin{equation}
\frac{i}{16\pi^3}[\frac{6}{a_{\psi}^3}g^4- \frac{1}{2^5 a_{\varphi}}\lambda^{2}]\frac{1}{d-6}.
\label{17}
\end{equation}
\begin{figure}[!h]
\centering
\includegraphics[scale=0.7]{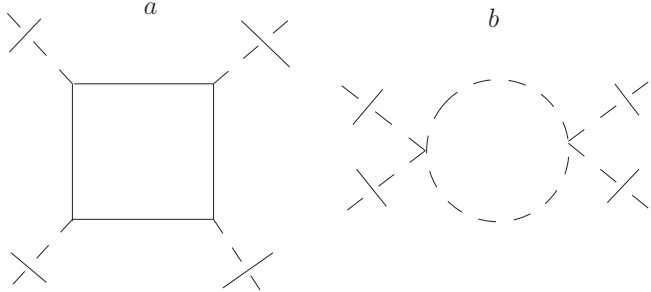}
\caption{Four point scalar functions.}
\label{E9}
\end{figure}
The vertex functions of this model satisfy the renormalization group equation

\begin{equation}
\left[\mu\frac{\partial}{\partial\mu}+D_{\varphi}+D_{\psi}+\beta_{g}\frac{\partial}{\partial g}+
\beta_{\lambda}\frac{\partial}{\partial\lambda}-N_B\gamma_{\varphi}-N_F\gamma_{\psi}\right]
\Gamma^{(N)}(p, m^2, b, a, \lambda, \mu)=0,
\label{5.46}
\end{equation}
where, we have defined the differential operators

\begin{equation}
D_{\varphi}\equiv\delta_{m^2}\frac{\partial}{\partial m^2}+\beta_{b_{\varphi}^{2}}\frac{\partial}{\partial b_{\varphi}^{2}}
+\beta_{a_{\varphi}^{2}}\frac{\partial}{\partial a_{\varphi}^{2}}
\label{5.46a}
\end{equation}
and

\begin{equation}
D_{\psi}\equiv\delta_{M}\frac{\partial}{\partial M}+\beta_{b_{\psi}}\frac{\partial}{\partial b_{\psi}}
+\beta_{a_{\psi}}\frac{\partial}{\partial a_{\psi}}.
\label{5.47}
\end{equation}

The parameter $\mu$ was introduced through the following replacements of the coupling constants
$g\rightarrow g \mu^{\frac{\epsilon}{2}}$ and $\lambda\rightarrow \lambda \mu^{\epsilon}$.

By substituting the one-loop expansions of the renormalized vertex functions we then get
the various {\it beta} functions (see the appendix \ref{apendiceb} for details), with the arguments indicating the order
of the coupling constant in which they have been calculated:

\begin{equation}
\beta_{b_{\varphi}^{2}}(g^2)=\frac{1}{32\pi^3}\frac{(b_{\varphi}^{2}+2 a_{\psi}M)}{a_{\psi}^3}g^2,
\label{M}
\end{equation}

\begin{equation}
\beta_{a_{\varphi}^{2}}(g^2)=\frac{1}{128\pi^3}\frac{(4 a_{\varphi}^{2}+ a_{\psi}^2)}{ a_{\psi}^3}g^2 \label{Y1},
\end{equation}

\begin{equation}
\beta_{b_{\psi}}(g^2)=\frac{b_{\psi}}{192\pi^3}
\frac{(a_{\varphi}+2 a_{\psi})}{a_{\varphi}a_{\psi}(a_{\varphi}+a_{\psi})^2}g^2,
\end{equation}

\begin{equation}
\beta_{a_{\psi}}(g^2)=\frac{1}{192\pi^3}
\frac{a_{\psi}(3a_{\varphi}+2a_{\psi})}{a_{\varphi}(a_{\varphi}+a_{\psi})^3}g^2\label{Y2},
\end{equation}

\begin{equation}
\beta_g(g^3)=\frac{1}{256\pi^3}
\left[\frac{2a_{\varphi}^3+4a_{\varphi}^2 a_{\psi}+3a_{\varphi}a_{\psi}^2+2 a_{\psi}^3}
{a_{\psi}^3 a_{\varphi}(a_{\varphi}+a_{\psi})^2}\right]g^3
\label{Y3}
\end{equation}

and
\begin{equation}
\beta_{\lambda}= \frac{3}{512\pi^3}\frac{\lambda^2}{a_{\varphi}^3}-\frac{3}{8\pi^3}\frac{g^4}{a_{\psi}^3}
+\frac{1}{16\pi^3}\frac{\lambda g^2}{a_{\psi}^3}.
\label{5.74}
\end{equation}
As the Eq. (\ref{M}) involves the mass of the fermion field, we need also the corresponding renormalization group function:
\begin{equation}
\delta_M(g^2)=\frac{[2a_{\varphi}a_{\psi}^2 b_{\varphi}^2+a_{\psi}^3b_{\varphi}^2+a_{\varphi}^3b_{\psi}^2
+2 a_{\varphi}^2a_{\psi}(b_{\psi}^2+2a_{\psi}M)]   }{256\pi^3 a_{\varphi}^3a_{\psi}(a_{\varphi}+a_{\psi})^2}g^2.
\end{equation}

The equations that govern the evolution of the effective Lorentz symmetry breaking parameters  may be separated in two sets. In the first one are  the equations for $\overline a_{\varphi}$, $\overline a_{\psi}$ and $\overline g$, which do not depend on the remaining parameters.
In the  second set are the  equations for the  other  parameters $\overline b_{\varphi}$ and $\overline b_{\psi}$, those need the input of the former set to be evaluated. Unfortunately these equations do not seem to have simple analytic solutions so that we  use numerical methods to investigate
their properties.
 Before proceeding, we need to stress what we mean by restoration of the Lorentz symmetry in this model. Obviously, we can never take
the symmetry breaking parameters $a_{\varphi}$ and $a_{\psi}$ equal to zero, since we would end up with a nonrenormalizable theory.
In this sense, we can not expect an exact Lorentz symmetry, which would correspond $a_{\varphi}=a_{\psi}=0$ and $b_{\varphi}=b_{\psi}$;
these are not even special points (for example, fixed points) of the above renormalization group equations.
However, we can consider the possibility of an approximate Lorentz symmetry arising in some specific low energy limit, depending
on a fine-tuning of the parameters involved. More precisely, we may find a region where the parameters
$\overline a_{\varphi}$ and $\overline a_{\psi}$ are sufficiently small, and, furthermore, the parameters $\overline b_{\varphi}$ and
$\overline b_{\psi}$ are
such that $\overline b_{\varphi}\approx \overline b_{\psi}$. Of course, this is not the ideal situation, but it could furnishes a positive
view concerning the  Lorentz symmetry  restoration in anisotropic field theories.
 To get a better insight on this possibility, we performed a numerical study, as described bellow.

Our results, obtained through the use of the numerical package of the Mathematica (and also checked with the Runge-Kutta 4th order method), show
that the fermionic parameters change much more slowly than the bosonic ones.
Actually, the changes of the tangents to the curves of some parameters are so small producing the impression that they are straight lines.
At higher momenta, the coupling constant $\overline g$, $\overline  a_{\varphi}$ and $\overline b_{\varphi}$ increase steeply for $t\sim 10^{9}$, indicating the existence of a singularity similar to the Landau pole  found  in many not asymptotically free field theories, see Figs. \ref{z1},
\ref{z2} and \ref{z5}. By contrast,  $\overline a_{\psi}$ and $\overline b_{\psi}$ increases slowly as shown in Fig. \ref{z1} and \ref{z2}.
For negative $t$, corresponding to the small momenta region, the general pattern is that all parameters  decrease, as remarked before, the parameters associated to the $\psi$ field do that  in a more slow rate than those associated to the $\varphi$ field
(see Figs. \ref{z3} and \ref{z4}). The graphical figures were drawn by taking the  initial value
$\overline g(0)=10^{-3}$ for the coupling constant and the initial  values of the other parameters all equal to one. Variations of these initial values do not qualitatively change the behavior of the effective parameters. It should be noticed  that below the lower ends of the curves the data are not reliable as the modulus of $t$ is very large whereas, at the same time, the effective parameters become
very small possibly generating large numerical errors.

 Thus, given a low energy region, in order to have an approximate Lorentz symmetric situation, one should modify the initial configuration of the
parameters $a_{\varphi}$ and $a_{\psi}$, so that
they will attain the desired range of values in the specified region. For example, as the $\overline a$'s monotonically decrease when $t$ is negative, this can be easily done by choosing the initial values of $a_{\varphi}$ and $a_{\psi}$ to be equal to the maximum values they should have in the region of interest.
Moreover, we need to adjust the initial configuration values
of the $b_{\varphi}$ and $b_{\psi}$ parameters, so that $\overline b_{\varphi}\approx \overline b_{\psi}$ in the same region. Notice that, these
two requirements can be satisfied due to the decoupling of the equations for $\overline a$'s and $\overline b$'s, as mentioned before.

\begin{figure}[!h]
\centering
\includegraphics[scale=0.75]{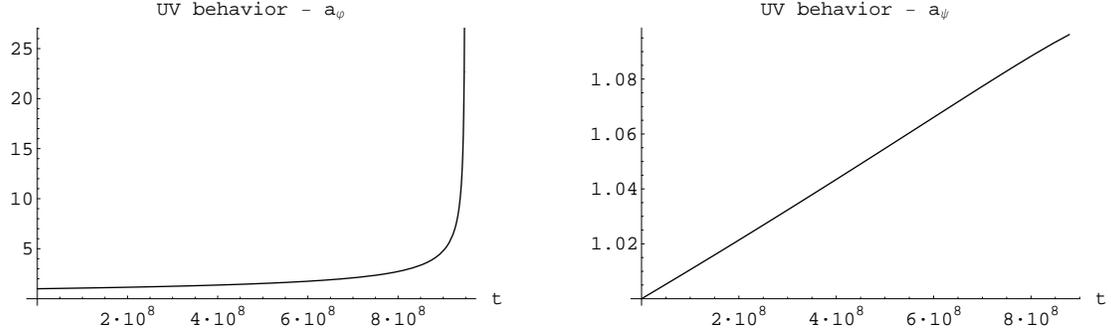}
\caption{The ultraviolet behaviors of the parameters $a_{\varphi}$ and $a_{\psi}$.}
\label{z1}
\end{figure}

\begin{figure}[!h]
\centering
\includegraphics[scale=0.75]{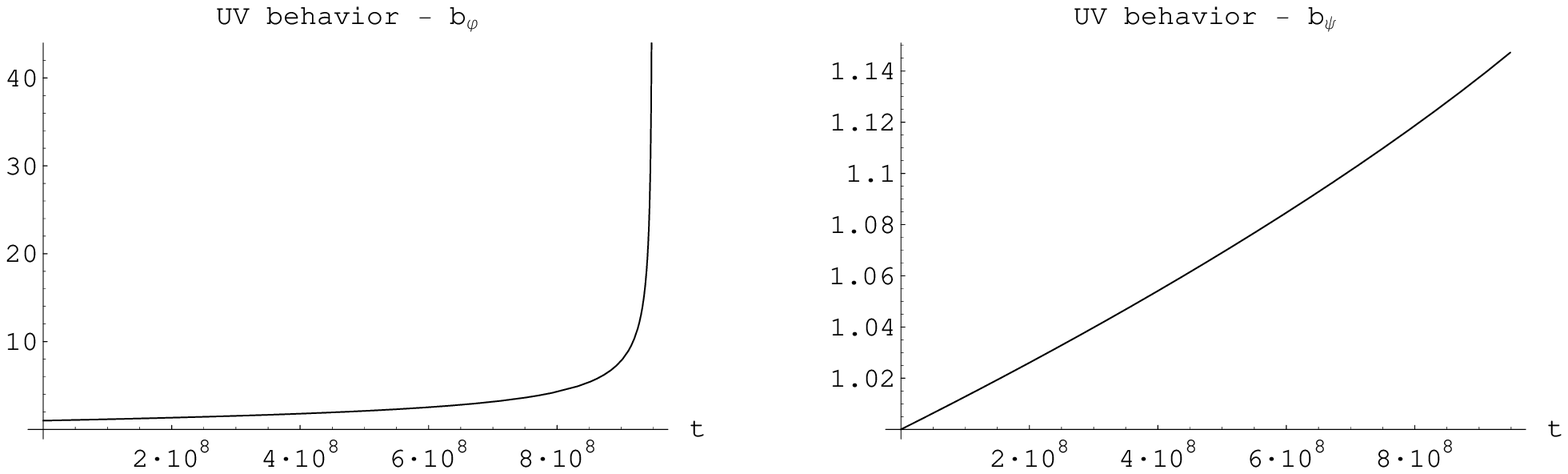}
\caption{The ultraviolet behaviors of the parameters $b_{\varphi}$ and $b_{\psi}$.}
\label{z2}
\end{figure}

\begin{figure}[!h]
\centering
\includegraphics[scale=0.75]{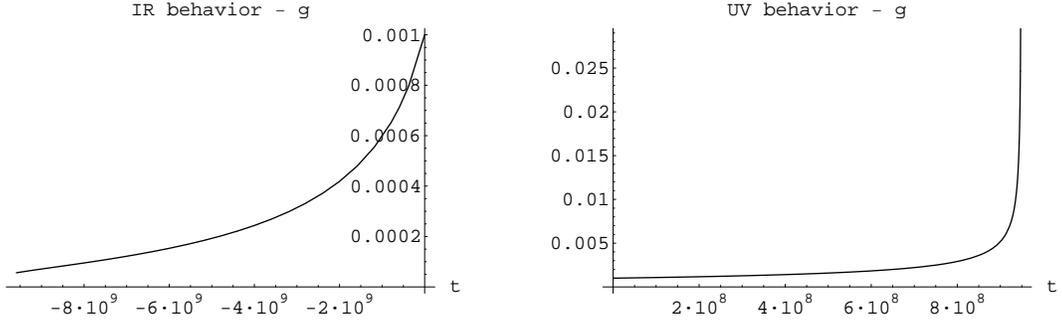}
\caption{The general behavior of the coupling constant g. The singularity in the UV region is similar to the Landau pole.}
\label{z5}
\end{figure}

\begin{figure}[!h]
\centering
\includegraphics[scale=0.75]{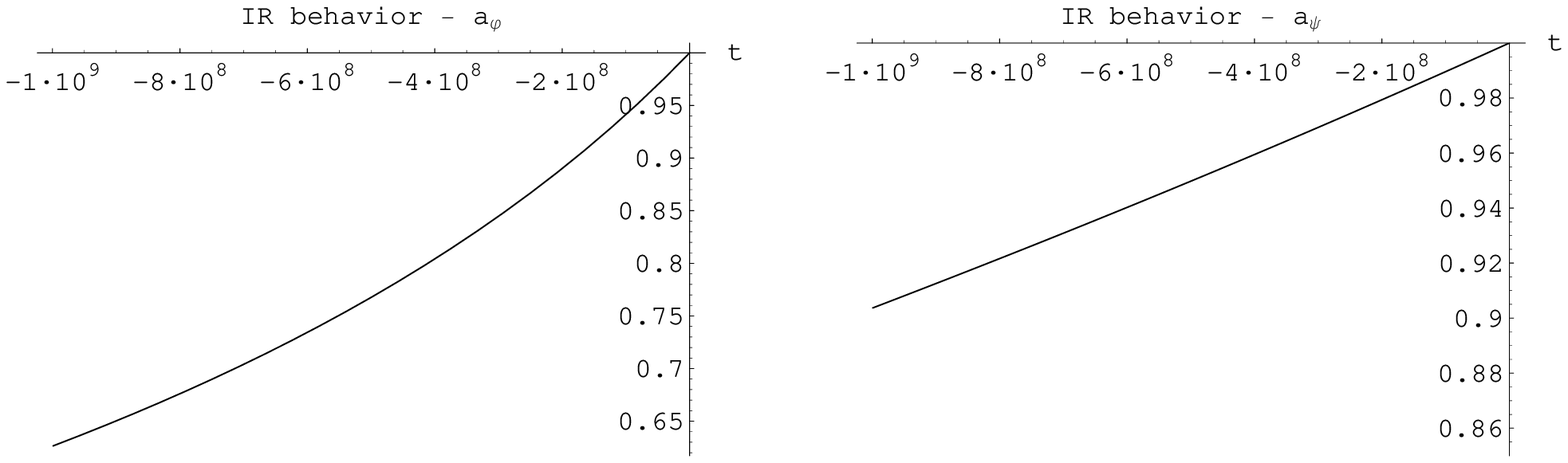}
\caption{The infrared behaviors of the parameters $a_{\varphi}$ and $a_{\psi}$.}
\label{z3}
\end{figure}

\begin{figure}[!h]
\centering
\includegraphics[scale=0.75]{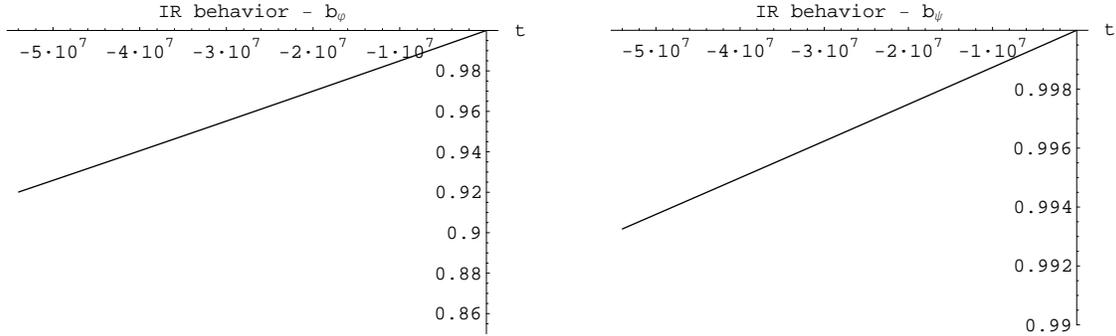}
\caption{The infrared behaviors of the parameters $b_{\varphi}$ and $b_{\psi}$.}
\label{z4}
\end{figure}

\section{Conclusions}

In this work we have examined some aspects  of anisotropic field theories as renormalization properties
and Lorentz symmetry restoration. In particular, we performed a renormalization group analysis of various models aiming to understand the infrared behavior of their effective parameters. Initially, we considered models without higher derivatives but with fields with different
light velocities; then in two specific instances we verified that the breaking of Lorentz invariance is soft, i.e., the Lorentz symmetry is recovered at low energies.

In the general situation where higher spatial derivatives are present, unless for super-renormalizable models, the idea that Lorentz symmetry is restored at low energies requires that the interactions be infrared stable and  some process of dimensional reduction, to cope with eventual
divergences appearing whenever the  higher derivative terms are eliminated. It should be pointed out  that, as may be straightforwardly derived from Eq. (\ref{1}),  the spatial dimensions in which the anisotropic, $d$, and Lorentz symmetric, $d_L$, versions of a given model without derivative couplings are renormalizable are  related by $d=z d_L$ where $z$ is the critical exponent which characterizes the anisotropy.

In the case of the $\varphi^{3}$ model, although the breaking parameters decrease with the energy, because of  the asymptotic freedom of the model,
the effective coupling increases and  no conclusion  can be achieved on the restoration of Lorentz symmetry within this perturbative approach.

Concerning the behavior of the parameters  under the renormalization group a more favorable situation occurs for the $\varphi^4$ and Yukawa models which are infrared stable. For the $\varphi^{4}$ model, we found a special analytical solution
in which the effective parameters increases monotonically from zero in the infrared to very large values at high momenta. For
the Yukawa model   the system of  equations that govern the changes of the effective parameters are intricate enough and no
closed analytical expression seems to be feasible. Our numerical solution  revealed that for high momenta a singularity, like the Landau pole, is present but for small momenta all Lorentz breaking parameters  decrease.  As discussed in the
text, our results indicate that we can find specific
low values of the energy scale where the approximate Lorentz symmetry may be achieved (in the sense of small $\overline a$'s and
$\overline b_{\varphi}\approx \overline b_{\psi}$), what nevertheless requires
a fine-tuning of the initial values the parameters.

In this study we have restricted ourselves to models with $z=2$. As we have seen, the investigation  of the Lorentz symmetry restoration
involves the determination of the infrared behavior of the Lorentz breaking parameters
and thus for models with higher values of $z$, which have  more parameters, the situation becomes more complex.
Another lateral remark concern models with
gauge symmetry; in this case, to keep the symmetry, covariant derivatives have to be used so that in the perturbative approach new interaction terms must to taken into account. For a discussion of the symmetries using the BPHZ approach on
anisotropic models, see the sequel \cite{Gomes1} of this work.

 We hope that  the ideas and methods we presented may be useful also in condensed matter physics,  in the contexts of quantum phase transitions and Lifshitz  models, i.e., in situations where the kind of  anisotropy here considered may be a natural concept \cite{Herbut,9b}.

\section{Acknowledgments}

The authors thank Dr. F. S. Bemfica for his assistance with the numerical analysis.
This work was partially supported by  Conselho
Nacional de Desenvolvimento Cient\'{\i}fico e Tecnol\'ogico (CNPq) and
 Funda\c{c}\~ao de Amparo a Pesquisa do Estado de S\~ao Paulo (FAPESP).

\appendix
\section{Integrals}\label{apendicea}
In this Appendix we will compute the pole part of some integrals that are relevant in our calculations. We begin by considering the following integral
\begin{equation}
J(x,y,z)\equiv \int \frac{dk_0}{2\pi}\frac{d^dk}{(2\pi)^d}\frac{k_0^x |{\bf k}|^y}
{[k_0^2-b^2{\bf k}^2-a^2 ({\bf k}^2)^2-m^2+i\epsilon]^z},
\label{4.19c}
\end{equation}
where $x,\, y$ and $z$ are such that the integral is  at most quartically divergent when the dimensional regularization is removed.

We use the representation
\begin{eqnarray}
J(x,y,z)&=& \frac{1}{i^z \Gamma(z)} \int \frac{dk_0}{2\pi}\frac{d^dk}{(2\pi)^d} k_0^x |{\bf k}|^y\int_{0}^{\infty} d \gamma \, \gamma^{z-1}{\rm e}^{i\gamma[k_0^2-b^2{\bf k}^2-a^2({\bf k}^2)^2-m^2+ i \epsilon]}\label{5}
 \nonumber\\
 &=& \frac{1}{i^z \Gamma(z)} \int_{0}^{\infty} d \gamma \, \gamma^{z-1}{\rm e}^{-i\gamma( m^2-i\epsilon)}\int\frac{dk_0}{2\pi} k_{0}^{x}{\rm e}^{i \gamma k_{0}^{2}}\int\frac{d^dk}{(2\pi)^d} |{\bf k}|^y{\rm e}^{i\gamma[-b^2 {\bf k}^2- a^2({\bf k}^2) ^2]}.
\end{eqnarray}

Now,
\begin{equation}
\int\frac{dk_0}{2\pi} k_{0}^{x}{\rm e}^{i \gamma k_{0}^{2}}= \frac{1}{4\pi}(1+(-1)^x) (-i\gamma)^{-\frac{(1+x)}2}\Gamma(\frac{x+1}2)
\end{equation}
and, denoting by $ \Omega_{d}=2\pi^{d/2}/\Gamma(d/2) $ the volume of the $ d $ dimensional unit sphere,
\begin{eqnarray}
\int\frac{d^dk}{(2\pi)^d} |{\bf k}|^y{\rm e}^{i\gamma[-b^2 {\bf k}^2- a^2({\bf k}^2) ^2]}&=& \frac{\Omega_d}{(2\pi)^d}\frac{(ia^2\gamma)^{-(d+y)/4}}{4}\left[ \Gamma(\frac{d+y}4)\, _1F_1\left (\frac{d+y}4,\frac12,\frac{i\gamma b^4}{4 a^2}\right)\right.\nonumber\\
&&- \frac{(i\gamma)^{1/2}}{|a|}b^2\Gamma(\frac{2+d+y)}4)\, _1F_1\left.\left (\frac{2+d+y)}4,\frac32,\frac{i\gamma b^4}{4 a^2}\right)\right ],\label{7}
\end{eqnarray}
where $ _1F_1 $ is the confluent hypergeometric function.
The divergence that appears in (\ref{5}) when the dimensional regularization is removed is due to the behavior of integrand for small $ \gamma $. Thus, to obtain its  pole part  we may use the approximations,
\begin{eqnarray}
_1F_1\left (\frac{d+y}4,\frac12,\frac{i\gamma b^4}{4 a^2}\right)&= &1 + i \frac{b^4(d+y)}{8 a^2} \gamma+ O(\gamma^2),
\nonumber\\ _1F_1\left (\frac{2+d+y)}4,\frac32,\frac{i\gamma b^4}{4 a^2}\right)&=&1+ O(\gamma).
\end{eqnarray}
Notice that in the last expansion we need to consider just the zeroth order
term because of the $ \gamma^{1/2} $ additional factor in (\ref{7}). All powers of $\gamma$ greater than those that we have considered produce finite results.
It is now straightforward to perform the remaining integrals and we obtain
\begin{eqnarray}
&&J(x,y,z)=\frac{1}{i^z\Gamma(z)}\frac{(1+(-1)^x)}{2}\Gamma\left(\frac{x+1}{2}\right)
\frac{a^{-(d+y+4)}}{(4\pi)^{\frac{d}2+1}\Gamma(\frac{d}{2})}
\nonumber\\
 &\times&\left[ a^4\Gamma\left(\frac{d+y}{4}\right)(-1)^{\frac{1}{8}(2-d-y)}
e^{\frac{i\pi x}{4}}(i m^2)^{\frac{1}{4}(2+d+2x+y-4z)}\Gamma\left(\frac{-2-d-2x-y+4z}{4}\right) \right.
\nonumber\\
&+&i\Gamma\left(\frac{d+y}{4}\right)\frac{b^4}{8}(d+y)(-1)^{\frac18(2-d-y)}
e^{\frac{i\pi x}{4}}(i m^2)^{\frac{1}{4}(-2+d+2x+y-4z)}
\nonumber\\&\times&\Gamma\left(\frac{2-d-2x-y+4z}{4}\right)
-i a^2 b^4\Gamma\left(\frac{d+y+2}{4}\right)(-1)^{\frac18(-d-14x-y)}e^{i 2\pi x}\nonumber\\
&\times&\left.(i m^2)^{\frac{1}{4}(d+2x+y-4z)}\Gamma\left(\frac{-d-2x-y+4z}{4}\right)\right].
\label{6}
\end{eqnarray}

Another typical integral that occurs in our calculation involves the product of propagators with different parameters as in

\begin{equation}
\int \frac{dk_0}{2\pi}\frac{d^dk}{(2\pi)^d}\frac{1}{(k_{0}^{2}-b^{2}_{1} {\bf k}^2-a^{2}_{1} ({\bf k}^2)^2-m^{2}_{1})^{z_1}} \frac{1}{(k_{0}^{2}-b^{2}_{2} {\bf k}^2-a^{2}_{2} ({\bf k}^2)^2-m^{2}_{2})^{z_2}}.
\end{equation}

In this case, we first join the two denominators by using Feynman trick
\begin{equation}
\frac{1}{A^{z_{1}}B^{z_{2}}}= \frac{\Gamma(z_{1}+z_{2})}{\Gamma(z_{1})\Gamma(z_{2})}\int_{0}^{1} dx\, \frac{x^{z_{1}-1}(1-x)^{z_{2}-1}}{[A x+ B(1-x)]^{z_{1}+z_{2}}}
\end{equation}
and then apply the formula (\ref{6}).

\section{Renormalization group parameters} \label{apendiceb}
Here we will present some details of the derivation of renormalization group parameters listed in the text. First notice that the boson and fermion two point vertex functions are given by
\begin{eqnarray}
\Gamma^{(2)}_{\varphi}(p)&=&i[p_0^2-b_{\varphi}{\bf p}^2-a_{\varphi}^{2} ({\bf p}^2)^2-m^2-g^2 (\text{Finite}_1-\ln\mu\text{Residue}_1)
\nonumber\\
&&-i\lambda(\text{Finite}_2-\ln\mu\text{Residue}_2)],
\label{5.47a}\\
\Gamma^{(2)}_{\psi}(p)&=&i[\gamma^0p^0-b_{\psi}\gamma\cdot{\bf p}-a_{\psi}{\bf p}^2-M+
g^2 (\text{Finite}_3-\ln\mu\text{Residue}_3)],
\label{5.50}
\end{eqnarray}
where the residues have the form
\begin{eqnarray}
\text{Residue}_1&=&A_1+A_2 p_0^2+A_3 {\bf p}^2+A_4({\bf p}^2)^2\qquad\text{and}\qquad \text{Residue}_2=\widetilde{A}_1,
\label{5.48}\\
\text{Residue}_3&=&B_1+B_2 \gamma^0 p^0+B_3 \gamma\cdot{\bf p}+B_4\,{\bf p}^2
\end{eqnarray}
and the $A$'s  and $B$'s  can be read directly from (\ref{9}) and (\ref{12}-\ref{15}).
Similarly,   the three point fermion-boson vertex function  and four point of the boson field have the  expressions

\begin{eqnarray}
\Gamma^{(3)}&=&-g\Gamma^5-g^3(\text{Finite}_4-\ln\mu\text{Residue}_4),
\label{5.52}\\
\Gamma^{(4)}&=&-i\lambda+g^4(\text{Finite}_5-\ln\mu\text{Residue}_5)-\lambda^2(\text{Finite}_6-\ln\mu\text{Residue}_6),
\label{5.54}
\end{eqnarray}
where $\text{Residue}_i$,  $i=4,5,6$, are given by
\begin{eqnarray}
\text{Residue}_4&=&C_1=\frac{1}{128\pi^3}\frac{1}{a_{\varphi} a_{\psi}^2+a_{\varphi}^{2} a_{\psi}}\Gamma^5, \qquad \text{Residue}_5=D_1=\frac{3i}{8\pi^3}\frac{1}{a_{\psi}^3},\\
\text{Residue}_6&=&\widetilde{D}_1=i\frac{3}{512 \pi^3}\frac{1}{a_{\varphi}^{3}}.
\end{eqnarray}
By replacing these expressions into (\ref{5.46}) and equating to zero the coefficient of each power of the coupling constant we get
\begin{equation}
\beta_{\lambda}(g^4)=i{D}_1 g^4, \qquad \beta_{\lambda}(\lambda^2)=-i\widetilde{D}_1\lambda^2
\label{5.67}
\end{equation}
and
\begin{equation}
\beta_{\lambda}(\lambda g^2)=4\lambda \gamma_{\varphi}(g^2),
\label{5.69}
\end{equation}
so that

\begin{equation}
\beta_{\lambda}=\frac{3}{512 \pi^3}\frac{\lambda^{2}}{a_{\varphi}^{3}}-\frac{3}{8\pi^3}\frac{g^{4}}{a_{\psi}^3}+\frac{1}{16\pi^3}\frac{\lambda g^{2}}{a_{\psi}^3}.\label{b1}
\end{equation}

We have also

\begin{equation}
\beta_{b^{2}_{\varphi}}(g^2)=-(b^{2}_{\varphi} A_2+A_3)g^2=\frac{1}{32\pi^3}\frac{({b^{2}_{\varphi}}+2a_{\psi}M)}{a_{\psi}^3}g^2,
\end{equation}

\begin{equation}
\beta_{{a^{2}_{\varphi}}}(g^2)=-(a^{2}_{\varphi} A_2+A_4)g^2=\frac{1}{128\pi^3}\frac{(4 {a^{2}_{\varphi}}+a_{\psi}^2)}{a_{\psi}^3}g^2,
\end{equation}

\begin{equation}
\beta_{b_{\psi}}(g^2)=-(b_{\psi} B_2+B_3)g^2=\frac{b_{\psi}}{192\pi^3}
\frac{(a_{\varphi}+2a_{\psi})}{a_{\varphi}a_{\psi}(a_{\varphi}+a_{\psi})^2}g^2,
\end{equation}

\begin{equation}
\beta_{a_{\psi}}(g^2)=-(a_{\psi} B_2+B_4)g^2=\frac{1}{192\pi^3}
\frac{a_{\psi}(3a_{\varphi}+2 a_{\psi})}{a_{\varphi}(a_{\varphi}+a_{\psi})^3}g^2
\end{equation}
and

\begin{equation}
\beta_g(g^3)=\left(-\frac{A_2}{2}-B_2+C_1\right)g^3=\frac{1}{256\pi^3}
\left[\frac{2a_{\varphi}^3+4a_{\varphi}^2 a_{\psi}+3a_{\varphi}a_{\psi}^2+2 a_{\psi}^3}
{a_{\psi}^3 a_{\varphi}(a_{\varphi}+a_{\psi})^2}\right]g^3.
\end{equation}


\end{document}